\documentclass[aps,prl,amsmath,amssymb,
superscriptaddress,
reprint]{revtex4-2}

\usepackage{graphicx}
\usepackage{dcolumn}
\usepackage{hyperref}
\usepackage{subfiles}
\usepackage[T1]{fontenc}
\usepackage{mathptmx}
\usepackage[dvipsnames]{xcolor}
\usepackage{etoolbox}

\begin{document}

\title{Opto-Electrical Detection of Donor Bound Excitons in Silicon-on-Insulator Substrate}

\author{A. Kanniainen}
\email{antti.j.kanniainen@jyu.fi}
\affiliation{Department of Physics and Nanoscience Center, University of Jyväskylä, Jyväskylä, Finland}

\author{A.S. Kumar}
\affiliation{Department of Physics and Nanoscience Center, University of Jyväskylä, Jyväskylä, Finland}

\author{A. Sammak}
\affiliation{QuTech and Netherlands Organisation for Applied Scientific Research (TNO), Delft, Netherlands}

\author{G. Scappucci}
\affiliation{QuTech and Kavli Institute of Nanoscience, Delft University of Technology, Delft, Netherlands}

\author{J.T. Muhonen}
\email{juha.t.muhonen@jyu.fi}
\affiliation{Department of Physics and Nanoscience Center, University of Jyväskylä, Jyväskylä, Finland}

\date{\today}

\begin{abstract}
Spin of a donor bound electron in silicon has been shown to be a very coherent qubit system but lacks a coherent optical interface. There does, however, exist a donor bound exciton transition that can be excited optically but decays dominantly via an Auger recombination producing an electrical signal. This provides an opto-electrical pathway for spin readout. Scaling this readout to single-spin level will require interfacing the spins with silicon photonics for efficient guiding of photons, which in turn will require moving to silicon-on-insulator (SOI) substrates. Here we demonstrate first ensemble opto-electronic measurements of donor bound excitons in SOI material, including isotopically purified $^{28}$Si device layers. The experiments show pronounced shifts in the resonance wavelengths and broadenings of the transition linewidths compared to the bulk experiments. 
\end{abstract}

\maketitle

\section{\label{sec:intro} Introduction}

Electron spins bound to phosphorus donors in silicon are among the most coherent qubit systems in the solid state. Single-atom spin qubits in silicon have achieved coherence times of the order of seconds~\cite{Muhonen2014}, gate fidelities surpassing 99.9\%~\cite{muhonen_quantifying_2015,Madzik2022}, and single-shot electrical readout~\cite{Morello2010}. These achievements, rooted in the original vision of Kane~\cite{Kane1998}, position donor spins as promising candidates for scalable quantum information processing~\cite{Morello2020}. It, however, still remains a challenge to scale the donor spin qubit architectures up to the scale required for fault tolerant quantum computing, partially due to lack of convenient and scalable readout methods.

The donor bound exciton (D$^0$X) transition provides a promising opto-electrical pathway for donor spin readout. In this process, a photon resonant with the D$^0 \rightarrow$ D$^0$X transition creates a bound exciton state consisting of the donor-bound electron plus an additional electron-hole pair. Since the transition energy depends on the initial spin state of the donor electron, the optical excitation is spin-selective~\cite{Yang2006,Sekiguchi2010} at non-zero magnetic fields. The exciton then decays predominantly via an Auger process, in which one electron recombines with the hole and the remaining electron is ejected into the conduction band, yielding a measurable change in photoconductivity. 
This hybrid opto-electrical detection scheme has been used in bulk experiments for some time~\cite{Yang2006,yang_simultaneous_2009,steger_quantum_2012, Ross2019}, but was demonstrated with local electrodes first by Lo~\emph{et al.}~\cite{Lo2015}. Subsequent experiments have studied the strain effects of the local electrodes to the transitions~\cite{Loippo2023,Conti2024}. These strain effects are well described by the Pikus-Bir Hamiltonian~\cite{PikusBir1974} and have been comprehensively characterized for P, As, and Sb donors under controlled uniaxial stress~\cite{mansir_linear_2018,Vogl2025}.

All prior D$^0$X detection experiments have been performed on bulk silicon substrates. However, scaling toward localized single-spin readout and integration with photonic circuits demands moving to silicon-on-insulator (SOI) substrates. The motivations for this are at least two-fold. First, SOI is the standard platform for silicon photonics, where thin device layers support single-mode waveguides and high-quality-factor optical resonators. Integrating donors into such structures could greatly enhance the efficiency of guiding excitation photons to the donor sites. One could also consider increasing the radiative decay channel via Purcell enhancement using photonic cavities, potentially enabling optical readout of single spins~\cite{Nur2019}. 
Second, SOI substrates naturally support the fabrication of suspended nanobeam optomechanical cavities, where mechanical resonators can be coupled to the donor spins, enabling new quantum hybrid systems that combine spins, phonons, and telecom-wavelength photons in a single integrated platform~\cite{shandilya_optomechanical_2021, Raniwala2025,Lyyra2025}.

In this work, we present the first opto-electronic detection of donor bound exciton transitions in SOI substrates by localized measurements of the laser-induced photoconductivity change at the D$^0$X resonance wavelength near 1078~nm. We investigate three SOI samples with different device layer thicknesses and phosphorus doping concentrations. Our experiments reveal D$^0$X resonances in all samples, but with pronounced shifts in the transition energy and substantial broadening of the linewidths compared to bulk silicon. In the thickest SOI sample (1~$\mu$m device layer), where donors were introduced during crystal growth, we resolve the heavy-hole and light-hole splitting and the magnetic field dependancy of the transitions, and achieve good agreement with the established strain model. In thinner samples (300~nm), to which donors where introduced by ion implantation, the broadening prevents resolving of the individual transitions in the magnetic fields used in our measurements (below 400 mT). 

\section{\label{sec:setup} Experiments}

\begin{figure}
    \centering
    \includegraphics[width=1\linewidth]{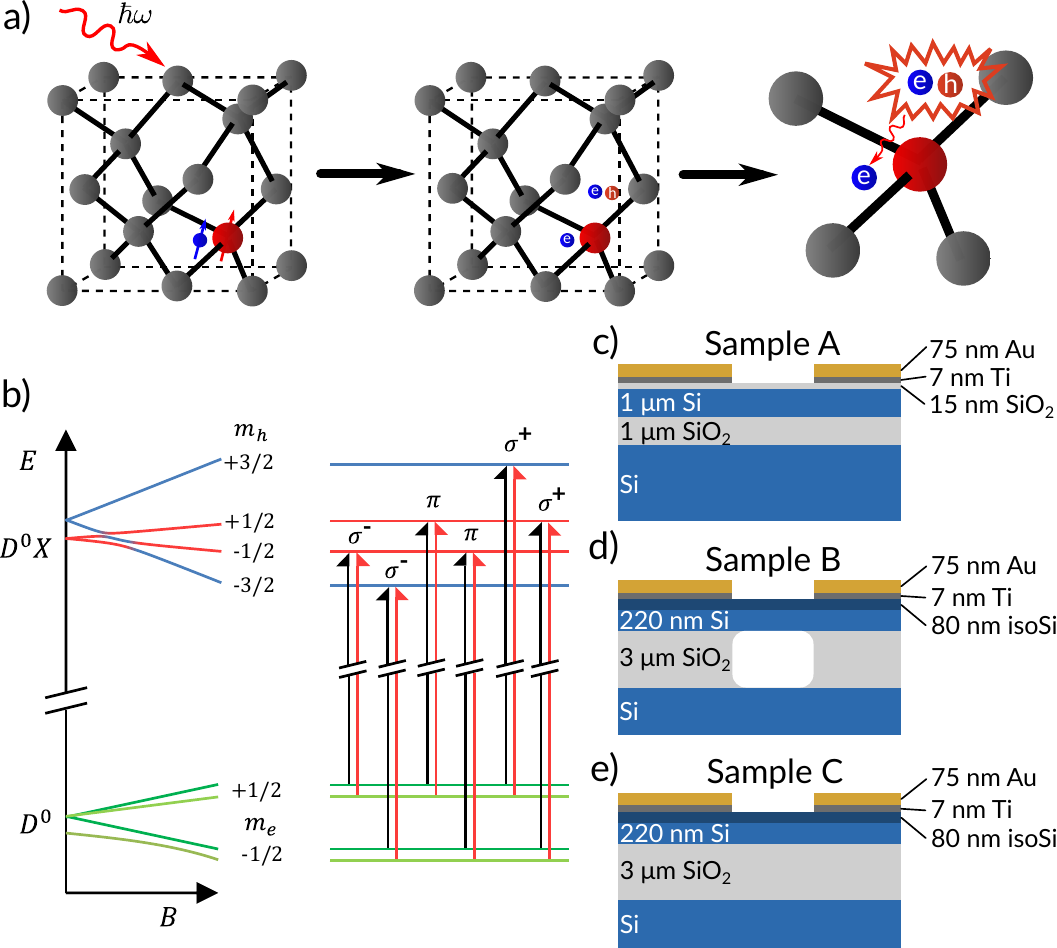}
    \caption{(a) Schematic illustration of the bound exciton recombination process at a neutral donor (D$^0$X) in a semiconductor lattice. A photon ($\hbar \omega$) creates an electron-hole pair bound to the donor site, which subsequently decays via an Auger process, ejecting a free electron into the conduction band. (b) Energy level diagram of the neutral donor bound exciton (D$^0$X) transition as a function of magnetic field B, showing the Zeeman splitting of the hole states ($m_h$ = $\pm$1/2, $\pm$3/2) and electron states ($m_e$ = $\pm$1/2), with allowed optical transitions indicated for $\sigma^-$, $\sigma^+$ and $\pi$ polarizations. (c) Cross-sectional schematic of Sample A, consisting of a 1 $\mu$m Si device layer on a 1 $\mu$m SiO$_2$ buried oxide, with a 15 nm SiO$_2$ gate dielectric and 7 nm Ti / 75 nm Au top contacts. (d) Cross-sectional schematic of Sample B, comprising of an 80 nm isotopically enriched silicon (isoSi) layer on a 220 nm Si device layer on a 3 $\mu$m SiO$_2$ buried oxide, with identical 7 nm Ti / 75 nm Au metallization. The silicon device layer was underetched between the electrodes to form a suspended membrane. e) Cross-sectional schematic of Sample C. The structure is identical to Sample B, except that the device layer was not suspended.}
    \label{fig:schema}
\end{figure}

The measurement principle and sample structures are illustrated in Fig.~\ref{fig:schema}. We performed measurements on three different SOI substrates. In sample~A the device layer thickness was 1~$\mu$m and the crystalline orientation was [100]. Phosphorus doping concentration of $5\times10^{15}$~cm$^{-3}$ was introduced during crystal growth. The device layer was grown by the Czochralski method, which results in a higher impurity concentration than float-zone silicon and, consequently, in broader exciton transition linewidths~\cite{Safonov1996}. The buried oxide layer (BOX) was 1~$\mu$m thick. Additionally, sample~A had approximately 15~nm of thermally grown oxide on top of the device layer, that was grown before the deposition of the readout electrodes. Sample~B had a thinner device layer of approximately 300~nm, consisting of 220~nm of natural silicon and an additional 80~nm layer of isotopically purified $^{28}$Si ($^{29}$Si concentration less than 800 ppm, measured with SIMS). The BOX layer was 3~$\mu$m thick. Phosphorus doping was introduced via ion implantation at a dose of $1\times10^{12}$~cm$^{-2}$ and an acceleration voltage of 20~keV. The implantation energy was chosen to place the peak phosphorus concentration centered within the isotopically purified layer, yielding an average dopant concentration of approximately $1\times10^{17}$~cm$^{-3}$ in the $^{28}$Si layer. Sample~C was structurally similar to sample~B, but had a higher ion implantation dose of $1\times10^{13}$~cm$^{-2}$, resulting in an average phosphorus concentration of approximately $1\times10^{18}$~cm$^{-3}$ within the isotopically purified layer.

\begin{table}[tb]
\caption{\label{tab:samples} Summary of SOI sample parameters. All samples have [100] crystalline orientation of the device layer.}
\begin{ruledtabular}
\begin{tabular}{lccc}
 & Sample A & Sample B & Sample C \\
\hline
Device layer thickness & 1~$\mu$m & 300~nm & 300~nm \\
$^{28}$Si layer thickness & --- & 80~nm & 80~nm \\
P concentration (cm$^{-3}$) & $5\times10^{15}$ & $1\times10^{17}$ & $1\times10^{18}$ \\
Doping method & in-growth & implantation & implantation \\
Surface oxide & 15~nm thermal & none & none \\
Suspended & No & Yes & No \\
\end{tabular}
\end{ruledtabular}
\end{table}

Microfabricated electrodes were defined on top of the device layers using PMMA resist and electron-beam lithography, followed by metal deposition via ultra-high-vacuum electron beam evaporation and a standard lift-off process. A metal stack consisting of 7~nm of titanium and 75~nm of gold was used for the electrodes. In sample~B, the device layer was suspended between the electrodes by underetching the buried oxide layer in hydrofluoric acid (HF), while samples~A and C remained fully supported on the buried oxide. The geometry of the electrodes was identical in all samples, with a 100~$\mu$m spacing between the electrodes.

A tunable diode laser (New Focus TLB-6700) with range 1045-1085~nm and specified linewidth <200~kHz is focused onto the region between the electrodes using a lens with NA 0.83, mounted above the sample, which enables the excitation of excitons within a confined area. Electronic transport characteristics were measured by either applying a DC bias and measuring the current through the sample or employing a lock-in amplifier to measure changes in the complex AC impedance (magnitude and phase of response). A static magnetic field was applied parallel to the sample surface. All measurements were performed at cryogenic temperatures ($T \approx 4$~K).

\section{\label{sec:results} Results and Discussion}

\begin{figure}
    \centering
    \includegraphics[width=\linewidth]{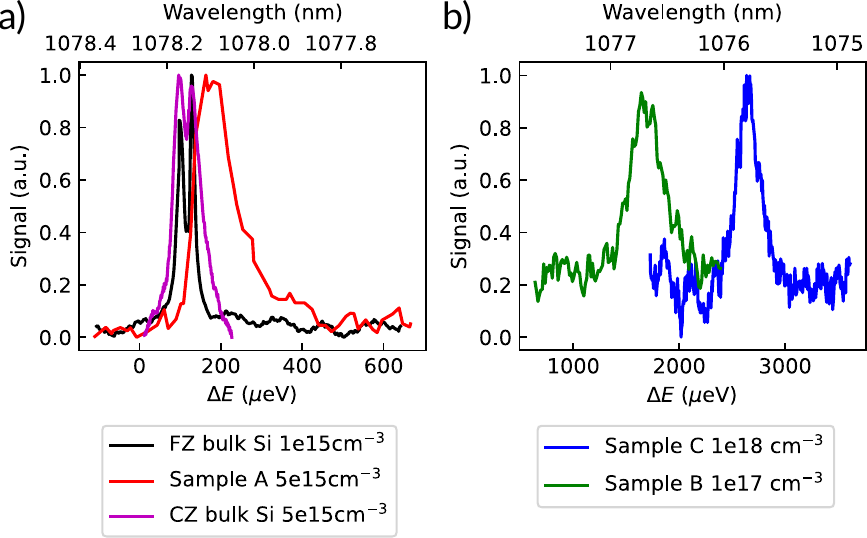}
    \caption{\label{fig:allpeaks} Electrically detected exciton spectra for silicon samples with varying phosphorus concentrations and substrates. (a) Comparison between float-zone (FZ) and Czochralski (CZ) bulk silicon (black and purple, respectively) \cite{Loippo2023} and SOI sample~A (red), showing the broadened and shifted exciton resonance peak in SOI. (b) Exciton spectra for samples~B and C on a wider energy scale, illustrating the further shift and stronger linewidth broadening at these thinner device layer samples with higher ion implanted doping concentrations.}
\end{figure}

Measurements on samples A, B and C all reveal opto-electrical resonances near the donor bound exciton (D$^0$X) transition wavelength, as shown in Fig.~\ref{fig:allpeaks}.
In sample~A, the resonance is shifted to higher energy by $\sim$100~$\mu$eV relative to the D$^0$X transition in bulk silicon, with a linewidth about twice as large as in the comparable bulk samples with similar electrodes measured in Ref.~\onlinecite{Loippo2023}. In samples~B and C, the resonances are shifted further, by $\sim$2~meV, and their linewidths are roughly an order of magnitude larger than in the bulk samples. Furthermore, the signal is weaker compared to the signal in sample~A, which can partly be explained by the thinner active volume. Although the volume doping concentration is higher in samples~B and C, the total number of optically active donors per unit area within the laser spot is comparable due to the much thinner device layer ($\sim$80~nm of doped $^{28}$Si versus 1~$\mu$m in sample~A). Due to the weaker signal, a periodic background arising from Fabry-Pérot interference in the handle wafer becomes visible in the spectra of sample~B, and is removed with a band-stop filter in the Fourier domain (see supplementary material~\cite{supplementary}).

Several factors intrinsic to the device geometry and the material stack of the samples~B and C can explain the shift of the exciton resonance to higher energies.
First, the reduced thickness of the silicon device layer is expected to enhance the strain fields arising from the thermal expansion mismatch between Si and the SiO$_2$ buried oxide~\cite{Iida1999}, as well as the metallic electrodes~\cite{Loippo2023, Conti2024}. Strain modifies the bound-exciton energies through the Pikus-Bir Hamiltonian~\cite{PikusBir1974}. In bulk, even the comparatively mild strain from microfabricated metal electrodes has been shown to produce measurable peak splittings and shifts of order 10-200~$\mu$eV~\cite{Loippo2023,Conti2024}. In SOI, the strain environment is expected to be significantly more severe, consistent with the much larger shifts of order 1--3~meV observed in samples~B and C, see modelling and discussion below.
Second, the ion implantation process used to introduce phosphorus donors in samples~B and C could leave residual lattice damage even after thermal annealing~\cite{Sumikura2011}, contributing additional shifts. 
Third, unintentional impurities such as carbon and oxygen that could be present at elevated concentrations near interfaces or within the isotopically purified layer could perturb the local environment of the donors and further modify the exciton binding energy.

\begin{figure*}
    \centering
    \includegraphics[width=0.75\linewidth]{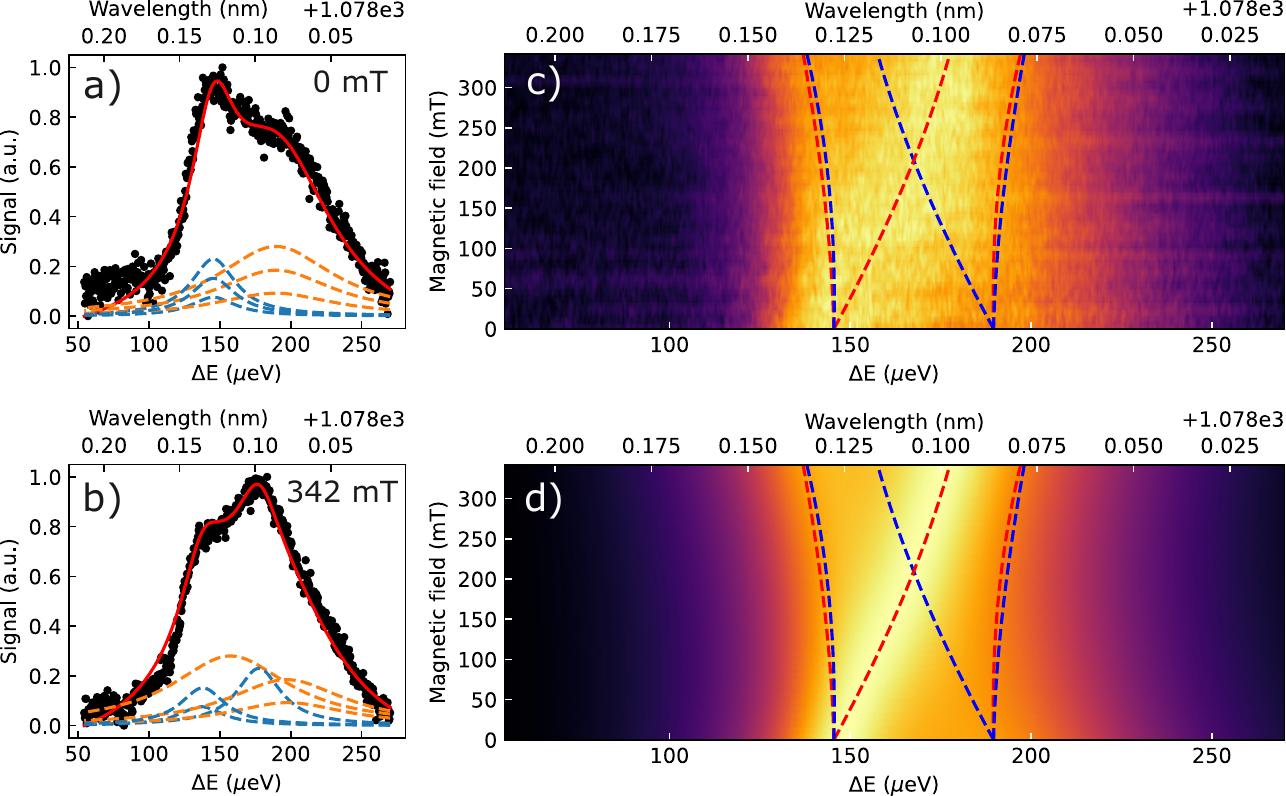}
    \caption{\label{fig:1umSOImagfield} (a,b) Phase response of the electrical signal as a function of laser detuning $\Delta E$ of sample~A at magnetic fields of 0~mT and 342~mT, respectively. The laser detuning is set to zero at the literature value for the phosphorus donor bound exciton transition in isotopically purified $^{28}$Si at~$\approx$1078.26~nm \cite{Yang2006}. Red curves represent the total fit, and dashed curves indicate individual Lorentzian components associated with the exciton transitions. Blue and orange dashed curves indicate the heavy-hole and light-hole transitions. (c) Measured phase response versus magnetic field and detuning of sample~A, revealing field-dependent splitting of the exciton resonances. (d) Corresponding magnetic field dependence from the fitted model. Dashed red (blue) lines represent the exciton transitions originating from the donor electron spin up (down) state.}
\end{figure*}

In sample~A, a high-resolution wavelength sweep near the resonance reveals two closely overlapping peaks as shown in Fig.~\ref{fig:1umSOImagfield}(a).
When a magnetic field is introduced, the profile of these overlapping peaks changes as shown in Fig.~\ref{fig:1umSOImagfield}(b). The magnetic field is in-plane and parallel to the [100] crystal direction. 

To model this data, we employ the same theoretical model as in Ref.~\onlinecite{Lo2015}, fixing the conduction band and valence band deformation potentials to literature values (see supplementary information~\cite{supplementary}). The strain parameters are treated as fitting variables to account for lattice distortions. We model each transition as a Lorentzian and assign the three heavy-hole transitions one transition amplitude and linewidth and the three light-hole transitions another. This reproduces the two peaks of unequal height and width observed at zero field.
We also assign empirical relative weights to the six transitions, labeled in order of increasing energy at high magnetic fields, so that the two central transitions carry a weight of 1, the second and fifth a weight of 2/3, and the outermost pair a weight of 1/3.
Both the unequal heavy-hole and light-hole amplitudes and linewidths, and the relative weights are consistent with earlier measurements, both from our own work~\cite{Loippo2023} and from other research groups~\cite{Vogl2025,Lo2015}. Incorporating these refinements enables the model to achieve good agreement with the measured data as shown in Figs.~\ref{fig:1umSOImagfield}(c) and \ref{fig:1umSOImagfield}(d).

\begin{figure}
    \centering
    \includegraphics[width=\linewidth]{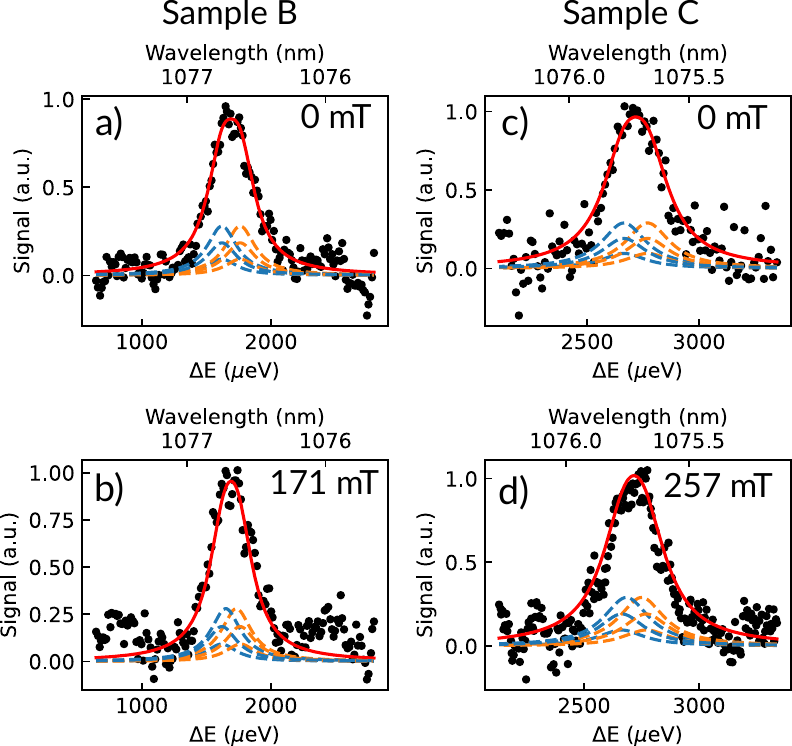}
    \caption{Measured spectra (black points) at zero field and at finite magnetic field for sample B (a, b) and sample C (c, d), together with the fitted model (red solid line) consisting of a sum of Lorentzian components (dashed lines). Blue and orange dashed curves indicate the heavy-hole and light-hole transitions. The applied magnetic field is indicated in each panel. The spectra are fitted using the maximum shear strain consistent with the data, yielding an upper limit on the shear strain and a lower limit on the linewidth. 
    }
    \label{fig:fig4}
\end{figure}

Figure~\ref{fig:fig4} shows the spectra of samples~B and C at zero field and in finite in-plane magnetic field. In contrast to sample~A, we do not observe a clear change in the magnetic field data. This is consistent with our model as the transitions shift by less than the observed linewidth over the accessible field range and remain unresolved, so the response is dominated by single broadened resonance whose shape shows only minimal change within our measurement accuracy. Because the response is a single broadened peak rather than resolved components, we cannot extract separate amplitudes for the heavy-hole and light-hole transitions, and we can only place limits on the off-diagonal strain terms and the linewidth, as discussed in supplementary information~\cite{supplementary}.

The fitted linewidths provide a measure of the inhomogeneous broadening in our samples. In sample~A we find linewidths of 19 and 50~$\mu$eV for the two peaks in zero field, comparable to previous on-chip electrical detection measurements in bulk natural silicon, where linewidths of approximately 25~$\mu$eV have been reported for Czochralski-grown material and 5-8~$\mu$eV for float-zone material~\cite{Loippo2023}. In samples~B and C the linewidths are considerably larger, 140~$\mu$eV in sample~B and 120~$\mu$eV in sample~C. Measurements in bulk isotopically purified $^{28}$Si show far narrower linewidths, both in electrical detection at low stress~\cite{Vogl2025} and in photoluminescence, where values down to 150~neV have been reported~\cite{Yang2006}. Since the device layers of samples~B and C are isotopically purified, isotope broadening cannot account for the observed linewidths, and we attribute the broadening primarily to inhomogeneous strain in the thin device layer. This is supported by the difference between the two samples, as the linewidths are largest in sample~B, where the device layer is suspended and the strain relaxation around the released membrane produces larger strain gradients than in sample~C, where the device layer remains attached to the buried oxide.

The strain parameters obtained from the fits allow a comparison of the strain environments in the three samples. In sample~A, where the narrower linewidths resolve the strain-induced splittings, the fits constrain both the diagonal and the off-diagonal strain components, yielding shear strains of $\varepsilon_{12} = \varepsilon_{13} = \varepsilon_{23} = 2.5\times10^{-6}$. In samples~B and C the linewidths are too large for the fits to be sensitive to the shear components, and equally good fits are obtained with the shear strain set to zero. The diagonal components are well constrained in all samples (assuming the zero-field energy shifts are solely due to strain) and differ substantially between them, with $\varepsilon_{11} = \varepsilon_{22} = \varepsilon_{33} = -6.1\times10^{-6}$ in sample~A, $-6.25\times10^{-5}$ in sample~B, and $-1.0\times10^{-4}$ in sample~C. The hydrostatic strain mainly shifts the transition energies, consistent with the resonances of samples~B and C appearing at higher energies than in bulk material. The compressive strain is largest in sample~C, where the device layer remains attached to the buried oxide, and is partially relaxed in the suspended device layer of sample~B.
We note that the extracted hydrostatic component depends on the assumed deformation potentials, whereas the comparison between the samples is unaffected as the same values are used throughout.

\section{\label{sec:level2} Summary and Conclusions}

We have demonstrated the first opto-electronic detection of phosphorus donor bound exciton transitions in silicon-on-insulator substrates, extending the hybrid optical-electrical readout technique~\cite{Lo2015} from bulk silicon to the SOI platform that underpins modern silicon photonics.

In sample~A (1~$\mu$m device layer, 5$\times$10$^{15}$~cm$^{-3}$ doping), the D$^0$X resonance is clearly resolved with moderate broadening compared to bulk CZ silicon, and the magnetic-field-dependent splitting of the exciton transitions can be well fitted using the Pikus-Bir Hamiltonian with strain parameters as fitting variables~\cite{Loippo2023,Vogl2025}. In samples~B and C (300~nm device layers, 10$^{17}$--10$^{18}$~cm$^{-3}$ doping via ion implantation into isotopically purified 80 nm $^{28}$Si layer), the resonances shift to substantially higher energies and exhibit significantly broader linewidths, precluding the resolution of individual exciton transitions with applied magnetic fields that were in our disposal (below 0.4 T). 

From a device perspective, the results are still encouraging: the D$^0$X transition is detectable in SOI, confirming that the Auger-mediated electrical readout mechanism remains functional even in thin device layers with substantial strain. In fact, this is the first demonstration that the mechanism can survive a shift of several meV:s. In conclusion, these first SOI measurements establish an important baseline for the ongoing development of donor-based quantum photonic and optomechanical platforms in silicon. Future work should focus on understanding and mitigating the linewidth broadening and exploring the behavior of D$^0$X transitions in fully processed photonic structures.


\begin{acknowledgments}
This project has received funding from the European Research Council (ERC) under the European Union’s Horizon 2020 research and innovation program (Grant Agreement No. 852428), from Research Council of Finland (Grant No. 321416), and European Innovation Council (EIC) (Grant Agreement No. 101185817). Funded by the European Union.
Views and opinions expressed are however those of the authors only and do not necessarily reflect those of the European Union or European Innovation Council or European Research Council. Neither the European Union nor the granting authorities can be held responsible for them.
\end{acknowledgments}

\bibliography{refs}

\clearpage
\onecolumngrid

\renewcommand{\thefigure}{S\arabic{figure}}
\renewcommand{\thetable}{S\arabic{table}}
\renewcommand{\theequation}{S\arabic{equation}}

\setcounter{figure}{0}
\setcounter{equation}{0}

\section{Supplementary Information}

\subsection{\label{app:model} Strain model fitting parameters}

The conduction band minimum in silicon is sixfold degenerate, however the donor potential breaks the translational symmetry of the crystal, lifting this degeneracy via valley-orbit coupling\cite{Luttinger1955}. Parameterizing the coupling by an overall binding energy $E_0$, and valley coupling constants $\Delta_1$ and $\Delta_2$ for neighboring and opposing valleys respectively, the valley-orbit Hamiltonian in the basis of the six individual conduction band valleys reads
\begin{equation}
    H_{VO} = 
    \begin{pmatrix}
    E_0&\Delta_1&\Delta_2&\Delta_2&\Delta_2&\Delta_2 \\
    \Delta_1&E_0&\Delta_2&\Delta_2&\Delta_2&\Delta_2 \\
    \Delta_2&\Delta_2&E_0&\Delta_1&\Delta_2&\Delta_2 \\
    \Delta_2&\Delta_2&\Delta_1&E_0&\Delta_2&\Delta_2 \\
    \Delta_2&\Delta_2&\Delta_2&\Delta_2&E_0&\Delta_1 \\
    \Delta_2&\Delta_2&\Delta_2&\Delta_2&\Delta_1&E_0 \\
    \end{pmatrix},
\end{equation}
Strain shifts the energies of the six valleys relative to one another, driving repopulation between them. By symmetry, this shift is fully described by a volume deformation potential $\Xi_d$ and a uniaxial deformation potential $\Xi_u$\cite{Karasyuk1992}. Adding these strain shifts to $H_{VO}$ gives the valley-repopulation Hamiltonian
\begin{equation}
    H_{VR}=H_{VO} + \Xi_d\sum_i \epsilon_{ii} + \Xi_u 
    \begin{pmatrix}
        \epsilon_{11}&&&&&0 \\
        &\epsilon_{11}&&&&  \\
        &&\epsilon_{22}&&&  \\
        &&&\epsilon_{22}&&  \\
        &&&&\epsilon_{33}&  \\
        0&&&&&\epsilon_{33}
    \end{pmatrix},
\end{equation}
where $\epsilon_{ii}$ are the diagonal components of the strain tensor.

The spin splitting of the neutral donor electron in a magnetic field $\mathbf{B}$ is described by the Zeeman Hamiltonian
\begin{equation}
    H_d^Z = \frac{1}{2}\mu_B g_d \mathbf{B}\cdot \boldsymbol{\sigma} , 
\end{equation}
where $\mu_B$ is the Bohr magneton, $g_d=1.9985$ is the donor electron $g$-factor, and $\boldsymbol{\sigma}$ are the Pauli matrices acting on the spin-1/2 electron state. The hyperfine interaction is neglected as it is unresolved in our experiments. The neutral donor state~D$^0$ energy is then obtained from the eigenvalues of the composite Hamiltonian $H_d^Z(\mathbf{B}) + H_{VR}(\epsilon)$.

The two electrons in the D$^0$X state form a spin-singlet, so the Zeeman response of the system is governed by the $J=3/2$ hole, described by the anisotropic $g$-factor model
\begin{equation}
    H_h^Z = \mu_B(g_1\mathbf{J}\cdot\mathbf{B} + g_2 \mathbf{J}^3\cdot \mathbf{B}),
\end{equation}
where $\mathbf{J}$ is the angular momentum vector operator for $J=3/2$, and $g_1$ and $g_2$ are the isotropic and anisotropic $g$-factors, respectively. The valence band maximum in unstrained silicon is fourfold degenerate, comprising the heavy-hole ($m_h=\pm 3/2$) and light-hole ($m_h=\pm 1/2$) doublets. Strain lifts this degeneracy and splits the two doublets in energy, which is captured by the Pikus-Bir Hamiltonian
\begin{equation}
    H_{PB}(\epsilon) = a\text{Tr}(\epsilon) +b\sum_{i}\biggl(J_i^2 - \frac{\textbf{J}^2}{3}\biggr)\epsilon_{ii} + \frac{d}{\sqrt{3}}\sum_{i\neq j} (J_i J_j + J_j J_i)\epsilon_{ij},
\end{equation}
where $a$,$b$ and $d$ are the hydrostatic, uniaxial and shear deformation potentials. The D$^0$X hole state energies are the eigenvalues of the total hole Hamiltonian $H_h^Z(\mathbf{B}) + H_{PB}(\epsilon)$.

The deformation potentials $b=-1.7\,\mathrm{eV}$ and $d=-5.1\,\mathrm{eV}$\cite{Lo2015} and the uniaxial valley-strain coupling $\Xi_u=17.41\,\mathrm{eV}$\cite{Vogl2025} are fixed to literature values, as is the combined parameter $a-~\Xi_d=-3.2\,\mathrm{eV}$\cite{Lo2015}, since $a$ and $\Xi_d$ cannot be distinguished within this model. The anisotropic hole $g$-factors are likewise fixed to $g_1=0.8$ and $g_2=0.24$\cite{kaminskil1980}. With these parameters fixed, the only free parameters in the fit are the strain tensor components. For sample A, we obtain $\epsilon_{11}=\epsilon_{22}=\epsilon_{33}=-6.1\times10^{-6}$ and $\epsilon_{12}=\epsilon_{13}=\epsilon_{23}=2.5\times10^{-6}$. Linewidths for the heavy-hole and light-hole transitions were $19~\mu$eV and $50~\mu$eV, and the ratio of the heavy-hole amplitude to light-hole amplitude was 1.22. For sample B, we obtain $\epsilon_{11}=\epsilon_{22}=\epsilon_{33}=-6.25\times10^{-5}$, and for sample C we obtain $\epsilon_{11}=\epsilon_{22}=\epsilon_{33}=-1\times10^{-4}$. The off-diagonal strain terms enter through the zero-field splitting of the transition, and in the fit they trade off directly against the intrinsic linewidth. With zero off-diagonal strain, the fit requires a somewhat larger linewidth to reproduce the observed peak width, while a nonzero off-diagonal strain contributes additional splitting that allows a correspondingly smaller linewidth to fit the same peak. Because the measured peak for samples B and C is a single broad feature rather than two resolved components, this trade-off cannot be broken, and the off-diagonal strain terms and the intrinsic linewidth cannot be determined independently. We can therefore only place an upper limit on the off-diagonal strain terms, above which the peak would resolve into two distinguishable components, and correspondingly a lower limit on the intrinsic linewidth. Likewise, the heavy-hole-to-light-hole amplitude ratio is unconstrained, as it does not noticeably affect the fit quality regardless of its value. For sample B, the lower limit of the linewidth is $140~\mu$eV and the upper limit to the off-diagonal components is $8\times10^{-6}$, and for sample C they are $120~\mu$eV and $6\times10^{-6}$, respectively. 

\subsection{\label{app:filtering} Band-stop filter in Fourier domain}

\begin{figure}
    \centering
    \includegraphics[width=0.8\linewidth]{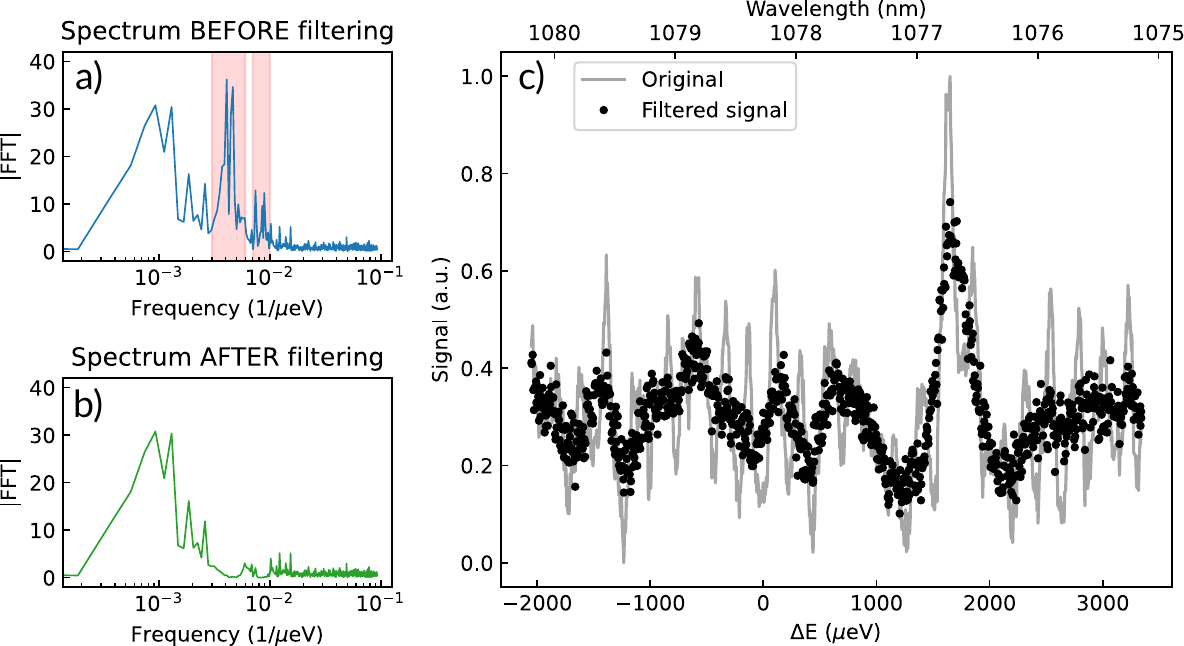}
    \caption{(a) Magnitude of the Fourier transform of a measured spectrum before filtering. The peak within the shaded region corresponds to the periodic modulation caused by interference in the silicon handle wafer. The band-stop filters suppress frequency components in the ranges 0.003-0.006 $\mu$eV$^{-1}$ and 0.007-0.010 $\mu$eV$^{-1}$, indicated by the shaded bands. (b) Fourier transform of the same spectrum after filtering, showing that only the interference component has been removed while the low-frequency content associated with the D$^0$X transitions and the high-frequency noise floor remain unaffected. (c) The spectrum before (gray line) and after (black points) filtering as a function of photon energy. The filtering removes the sinusoidal modulation of the background without altering the transition line shapes.}
    \label{fig:S1}
\end{figure}

The spectra of sample B exhibit a weak periodic background superimposed on the D$^0$X transitions, appearing in the Fourier domain as a distinct peak at a frequency of 0.0045 $\mu$eV$^{-1}$ together with its second harmonic at 0.009 $\mu$eV$^{-1}$. We attribute this modulation to Fabry-Pérot interference in the silicon handle wafer. At cryogenic temperatures, the band gap shifts so that the substrate is largely transparent in the measurement range, and light reflected from the wafer backside interferes with that reflected at the top surface. The measured interference frequency corresponds to a silicon layer thickness of approximately 730~$\mu$m (using $n_g \approx 3.8$, the group index of silicon near the band edge), in agreement with the nominal 725~$\mu$m thickness of the handle wafer. The presence of the second harmonic is consistent with the non-sinusoidal transmission function of a Fabry-Pérot cavity. Because the interference is periodic in energy with a fixed, independently verified frequency, it can be cleanly separated from the spectral features of interest. We therefore remove the modulation with narrow band-stop filters applied in the Fourier domain, suppressing only the components at the interference frequency and its second harmonic while leaving the D$^0$X line shapes unaffected as shown in Fig~\ref{fig:S1}. A similar periodic background is faintly visible in the spectra of sample C, but because of the higher doping concentration the signal is stronger relative to the modulation and the interference does not affect the analysis.

\end{document}